\documentstyle[aps,multicol,epsfig,epsf,psfrag]{revtex}

\draft
\newcommand{\beq}{\begin{equation}}
\newcommand{\eeq}{\end{equation}}
\newcommand{\bdis}{\begin{displaymath}}
\newcommand{\edis}{\end{displaymath}}
\newcommand{\bea}{\begin{eqnarray}}
\newcommand{\eea}{\end{eqnarray}}
\newcommand{\barr}{\begin{array}}
\newcommand{\earr}{\end{array}}

\begin{document}

\title{Creep rupture of viscoelastic fiber bundles}

\author{Raul Cruz Hidalgo$^{1}$, Ferenc
  Kun$^{1,2}$\footnote{Electronic
address:feri@ica1.uni-stuttgart.de},
and Hans. J. Herrmann$^{1}$} 

\address{$^1$Institute for Computational Physics, University of
Stuttgart,    
Pfaffenwaldring 27, 70569 Stuttgart, Germany\\
$^2$Department of Theoretical Physics, University of Debrecen, \\ 
P.O.Box: 5, H-4010 Debrecen, Hungary}

\date{\today}
\maketitle
\begin{abstract} 
We study the creep rupture of bundles of viscoelastic fibers
occurring under uniaxial constant tensile loading. A novel fiber bundle
model is introduced which combines the viscoelastic constitutive behaviour 
and the strain controlled breaking of fibers. 
Analytical and numerical calculations showed that above a
critical external load the deformation of the system monotonically
increases in time resulting in global failure at a finite time $t_f$,
while below the critical load the deformation tends to a constant
value giving rise to an infinite lifetime.
Our studies revealed that the nature of the transition between the two 
regimes, {\it i.e} the behaviour of  $t_f$ at the critical load $\sigma_c$,
strongly depends on the range of load sharing: for global load sharing 
$t_f$ has a 
power law divergence at $\sigma_c$ with a universal exponent of
$0.5$, however, for local load
sharing the transition becomes abrupt: at the critical load $t_f$
jumps to a finite value, analogous to second and first order phase
transitions, respectively. The acoustic response of the bundle during
creep is also studied. 
\end{abstract}

\pacs{PACS number(s): 46.50.+a, 62.20.Mk}

\begin{multicols}{2}
\narrowtext

Fiber reinforced composites are of great technological
importance due to their very good performance under extreme
circumstances. Under high steady stresses these fiber composites
may exhibit time dependent failure called creep rupture, which limits
their life time and consequently has a high impact on the
applicability of these materials for construction elements. Both
natural fiber composites like wood 
\cite{laufenberg,pu} and 
various types of fiber reinforced composites \cite{chiao,otani,farq}
show creep 
rupture phenomena, which have attracted continuous theoretical
and experimental interest over the past years. The underlying
microscopic failure mechanism of creep   
rupture is very complex depending on several characteristics of
the specific types of materials, and is far from being well
understood. From technological point of view, one of the most
important aspects of creep rupture is the 
statistics of life 
time (or time to failure) as a function of the external steady load, however,
only a limited number of systematic experimental works is available for
fiber reinforced composites
\cite{chiao,otani,farq}, more information has been
accumulated about natural fiber composites \cite{laufenberg,pu}.
In Ref.\ \cite{phoenix} a theoretical model of creep rupture of
brittle matrix composites reinforced with time dependent fibers was
worked out, where the fibers are assumed to have a finite life time
under a constant load in the spirit of the classical model of Coleman
\cite{coleman}. For natural fiber composites a so-called damage
accumulation model has been developed, which simply assumes
that the time derivative of the accumulated damage depends
exponentially on the external load history of the specimen
\cite{gerhards}.   

In the present letter we study the creep rupture of fiber composites
where the fibers have viscoelastic behaviour and the microscopic
damage mechanism leading to creep rupture is the strain dependent
breaking of fibers under the time evolution of the deformation of the
system.  
Creep failure tests are usually
performed under uniaxial tensile loading when the
specimen is subjected either 
to a constant load $\sigma_{\rm o}$ or to an increasing load
(ramp-loading) and the time evolution of the 
damage process is followed by recording the strain $\varepsilon$ of
the specimen and
the acoustic signals emitted by microscopic failure events.
In the present study we focus on the general aspects of creep rupture,
{\it i.e.} the behaviour of the life  
time of the bundle as a function of the external load, its dependence
on the range of load redistribution, furthermore, general aspects of
the acoustic response of the bundle are considered without fitting the 
theoretical results to any specific materials.

In order to work out a theoretical description of creep failure of
viscoelastic fiber composites we improve the 
classical fiber bundle model
\cite{coleman,daniels}
which has proven very successful in the 
study of fracture of disordered materials
\cite{hansen,phoenix1,leath,zhang,delapl,sornette1,sornette2,sornette3,kun2,yamir}.
Our model consists of $N$ parallel fibers having
viscoelastic constitutive behaviour. 
For simplicity, the pure viscoelastic behaviour of fibers is
modeled by a Kelvin-Voigt element which consists of a spring and a
dashpot in parallel 
and results in the constitutive equation
\begin{eqnarray}
  \label{eq:visco}
  \sigma_{\rm o} = \beta \dot{\varepsilon} + E\varepsilon, 
\end{eqnarray}
where $\beta$ denotes the damping coefficient, and  $E$ the Young
modulus of fibers, respectively.  
 Eq.\ (\ref{eq:visco}) provides
the time dependent deformation $\varepsilon(t)$ of a fiber at a
fixed external load  $\sigma_{\rm o}$
\begin{eqnarray}
  \label{solve}
  \varepsilon(t) = \frac{\sigma_o}{E}\left[1-e^{-Et/\beta} \right] +
  \varepsilon_{\rm o} e^{-Et/\beta},
\end{eqnarray}
where $\varepsilon_o$ denotes the initial strain at $t=0$. It can be
seen that  $\varepsilon(t)$ converges to $\sigma_{\rm o}/E$ for $t 
\to \infty$, which implies that the asymptotic strain fulfills Hook's
law. 

If no fiber failure occurs Eq.\ (\ref{solve}) would fully describe the
time evolution of the system. Motivated by the experimental
observations of the acoustic response \cite {acoust} of fiber
composites during creep, 
we introduce a strain controlled failure criterion 
to incorporate damage in the model: a fiber fails
during the time evolution of the system if
its strain exceeds a damage threshold $\varepsilon_d$, which is an
independent identically distributed random variable of fibers with
probability density 
$p(\varepsilon_d)$ and cumulative distribution ${\displaystyle 
  P(\varepsilon_d) = \int_0^{\varepsilon_d} p(x) dx}$. Similar
strain controlled breaking was recently used in Ref.\ \cite{menezes}.  
Due to the validity of Hook's law for the asymptotic strain values, the
formulation of the failure criterion in terms of strain instead of
stress implies that under a certain steady load the same amount of
damage occurs  as in the case of stress controlled failure, however,
the breaking of fibers is not instantaneous but distributed over
time. When a fiber fails its load has to
be redistributed to the 
intact fibers. As the simplest approach, we assume global load
sharing \cite{hansen,sornette1,sornette2,sornette3,kun2,yamir}, {\it
    i.e.} after a failure event the excess load is equally  
distributed among the intact fibers, and hence, at a certain strain
$\varepsilon$ the load on the surviving
fibers of the number  $N_s(\varepsilon)$ can be cast into the form 
$  \sigma(\varepsilon) = \sigma_{\rm o}N/N_s(\varepsilon) =
  \sigma_{\rm o}/(1-P(\varepsilon))$.
The time evolution of the system under a steady external load
$\sigma_o$ is finally described by the equation 
\begin{eqnarray}
  \label{eq:eom}
  \frac{\sigma_{\rm o}}{1-P(\varepsilon)} = \beta \dot{\varepsilon}
  +E\varepsilon, 
\end{eqnarray}
where the viscoelastic behaviour of fibers is coupled with the failure 
of fibers in a global load sharing framework.

For the behaviour of the solutions of Eq.\ (\ref{eq:eom}) two distinct
regimes can be distinguished depending on the value of the external
load $\sigma_{\rm o}$: When $\sigma_{\rm o}$ is below a critical value
$\sigma_{\rm c}$ Eq.\ (\ref{eq:eom}) has a stationary solution
$\varepsilon_s$, which can be obtained by setting $\dot{\varepsilon}=0$
\begin{eqnarray}
  \label{eq:stationary}
  \sigma_{\rm o} = E\varepsilon_s[1-P(\varepsilon_s)].
\end{eqnarray}
It means that until this equation can be solved for $\varepsilon_s$ 
at a given external load $\sigma_{\rm o}$, the solution  $\varepsilon(t)$ of
Eq.\ (\ref{eq:eom}) converges to 
$\varepsilon_s$  when $t\to \infty$, and no macroscopic failure
occurs. However, when 
$\sigma_{\rm o}$  exceeds the critical value $\sigma_{c}$ no stationary
solution exists, furthermore, $\dot{\varepsilon}$ remains always
positive, which implies that for $\sigma > \sigma_{c}$ the strain of the
system   $\varepsilon(t)$ monotonically increases until the 
system fails globally at a time $t_f$. 

In the regime $\sigma_{\rm o} \leq \sigma_{\rm c}$ Eq.\
(\ref{eq:stationary}) also provides the asymptotic constitutive
behaviour of the 
fiber bundle which can be measured by controlling the external load
$\sigma_{\rm o}$ and letting the system relax to $\varepsilon_s$. It follows 
from the above argument that the critical value of the load
$\sigma_{c}$ is the static fracture strength of the bundle which can be
determined from Eq.\ (\ref{eq:stationary}) as 
 $ \sigma_{c} = E\varepsilon_{c}[1-P(\varepsilon_{c})]$,
where $\varepsilon_{c}$ is the solution of the equation
$\displaystyle{\left. d\sigma_{\rm o}/d\varepsilon_s
  \right|_{\varepsilon_{\rm c}} = 0}$, as shown by Sornette
\cite{sornette1}.    
Since $\sigma_{\rm o}(\varepsilon_{\rm s})$ has a maximum of the value
$\sigma_{\rm c}$ at 
$\varepsilon_{\rm c}$, in the vicinity of $\varepsilon_{\rm c}$ it can be
approximated as  
\begin{eqnarray}
  \label{eq:series}
  \sigma_{\rm o} \approx \sigma_{\rm c} -A(\varepsilon_{\rm
    c}-\varepsilon_s)^2, 
\end{eqnarray}
where the multiplication factor $A$ depends on the probability
distribution $P$.
A complete description of the system can be obtained by solving the 
differential equation Eq.\ (\ref{eq:eom}). After separation of
variables 
the integral arises 
\begin{eqnarray}
  \label{eq:integ}
t =  \beta \int d\varepsilon 
  \frac{1-P(\varepsilon)}{\sigma_{\rm o}-E\varepsilon \left[
    1- P(\varepsilon)\right]} + C,
\end{eqnarray}
where the integration constant $C$ is determined by the initial
condition $\varepsilon(t=0)=0$.

The creep rupture of the viscoelastic bundle can be interpreted so
that for $\sigma_{\rm 
  o} \leq \sigma_{\rm c}$ the life time (or the time to failure) of
the bundle is infinite  
$t_f = \infty$, while above the critical load
$\sigma_{\rm o} > \sigma_{\rm c}$ global failure occurs at a finite
time $t_f$, which can be determined by
evaluating the integral Eq.\ (\ref{eq:integ}) over the whole domain of
definition of 
$P(\varepsilon)$.  From the theoretical and experimental point of view it
is very important how $t_f$ depends on the external load
above $\sigma_{\rm c}$. When $\sigma_{\rm o}$ is in the
vicinity of 
$\sigma_{\rm c}$, {\it i.e.} $\sigma_{\rm o}=\sigma_{\rm c}+\Delta
\sigma_{\rm o}$, where 
$\Delta \sigma_{\rm o} << \sigma_{\rm c}$, it can be expected that the 
curve of $\varepsilon(t)$ falls very close to $\varepsilon_{\rm c}$ for a very
long time and the breaking of the system occurs suddenly.
Hence, the total time to failure,
{\it i.e.} the  
integral in Eq.\ (\ref{eq:integ}), is dominated by the 
region close to $\varepsilon_{\rm c}$ when $\Delta \sigma_{\rm o}$ is
small. Making 
use of the power series expansion Eq.\ (\ref{eq:series}) the
integral in Eq.\ (\ref{eq:integ}) can be rewritten as
\begin{eqnarray}
  \label{eq:time}
  t_f \approx  \beta \int d\varepsilon 
  \frac{1-P(\varepsilon)}{\Delta \sigma_{\rm o} - A(\varepsilon_{\rm c}-\varepsilon)^2}, 
\end{eqnarray}
which has to be evaluated over a small $\varepsilon$
interval in the vicinity of $\varepsilon_{\rm c}$. After 
performing the integration 
it follows 
\begin{eqnarray}
  \label{critical}
  t_f \approx (\sigma_{\rm o} - \sigma_{\rm c})^{-1/2}, \qquad \mbox{for} \qquad
  \sigma_{\rm o} > \sigma_{\rm c}. 
\end{eqnarray}
Thus, $t_f$ has a power law divergence at $\sigma_{\rm c}$ with
a universal exponent $-\frac{1}{2}$ independent of the
specific form of the disorder  distribution $p(\varepsilon)$.

\begin{figure} 
\begin{center}
\psfrag{aa}{{\large $t_f$}}
\psfrag{bb}{{\large $Et/\beta$}}
\psfrag{cc}{{\Large $\varepsilon/\varepsilon_m$}}
\psfrag{dd}{{\Large $\varepsilon_{\rm c}$}}
\epsfig{bbllx=5,bblly=0,bburx=200,bbury=175, file=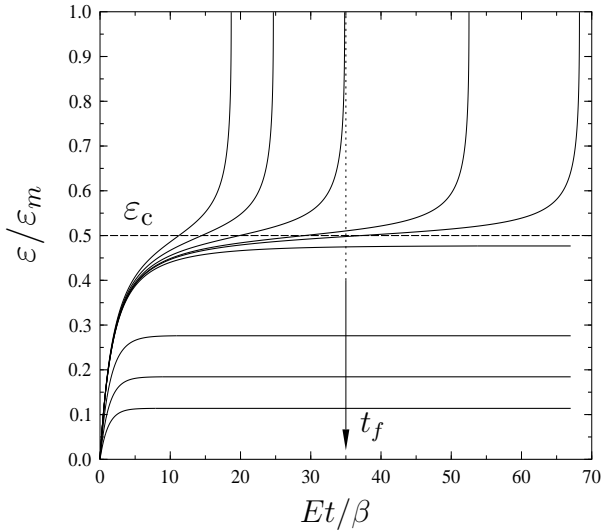, width=9cm}
\caption{The analytic solution $\varepsilon(t)$ given by Eqs.\
  (\ref{eq:below},\ref{eq:above}) 
    for several values of $\sigma_{\rm o}$ below and above
    $\sigma_{\rm c}$. The critical strain $\varepsilon_{\rm c}$ and the time to
    failure $t_f$ for one example are indicated.}
\label{fig:eps}
\end{center}
\end{figure}

For the purpose of explicit calculations we considered the case of
a uniform distribution of the damage thresholds between $0$ and a
maximum value $\varepsilon_m$, thus,
$p(\varepsilon_d)=1/\varepsilon_m$ and
$P(\varepsilon_d)=\varepsilon_d/\varepsilon_m$. The stationary
solution, the critical load and the corresponding critical strain can
be obtained as 
$\sigma_{\rm o} = E\varepsilon[1-\varepsilon/\varepsilon_m]$,
$\sigma_{c}=E\varepsilon_m/4$,
$\varepsilon_{c}=\varepsilon_m/2$, respectively. Finally, the solution 
of the integral Eq.\
(\ref{eq:integ}) 
 taking the initial condition also into account can be
cast into the implicit form 
\begin{eqnarray}
  \label{eq:below}
t= -\frac{\beta}{2E} && \left\{ 
  \frac{1}{\sqrt{1-\frac{4\sigma_{\rm o}}{E\varepsilon_m}}}
  \ln
  \frac{\frac{\varepsilon}{\varepsilon_m}
  \left[1+\sqrt{1-\frac{4\sigma_{\rm o}}{E\varepsilon_m}} \right]
  +\frac{2\sigma_{\rm o}}{E\varepsilon_m}}
  {\frac{\varepsilon}{\varepsilon_m}  \nonumber
   \left[1-\sqrt{1-\frac{4\sigma_{\rm o}}{E\varepsilon_m}} \right] 
 +\frac{2\sigma_{\rm o}}{E\varepsilon_m}} \right. \\ 
 && \left.  - \ln\frac{E\varepsilon^2-E\varepsilon_m\varepsilon+
      \sigma_{\rm o}\varepsilon_m}{\sigma_{\rm o}\varepsilon_m}
 \right\} 
\end{eqnarray}
for $\sigma_{\rm o} < \sigma_{c}$ (below the {\it critical point}), and 

\begin{eqnarray}
  \label{eq:above}
&&t=\frac{\beta}{E}\left\{
\frac{1}{\sqrt{\frac{4\sigma_{\rm o}}{E\varepsilon_m}-1}} 
  \left[   arc tg \frac{2\frac{\varepsilon}{\varepsilon_m}
        -1}{\sqrt{\frac{4\sigma_{\rm o}}{E\varepsilon_m}-1}} 
    \right. \right. \\
  && \left. \left. -
      arctg\frac{-1}{\sqrt{\frac{4\sigma_{\rm o}}{E\varepsilon_m}-1}}
    \right] 
     -\frac{1}{2}\ln \frac{E\varepsilon^2-E\varepsilon_m \varepsilon
      +\sigma_{\rm o} \varepsilon_m}{\sigma_{\rm o}\varepsilon_m} \right\}           \nonumber
\end{eqnarray}
for  $\sigma_{\rm o} > \sigma_{c}$ (above the {\it critical point}).
The behaviour of this analytic solution is illustrated
in Fig.\ \ref{fig:eps} for several different values of $\sigma_{\rm o}$.

The time to failure $t_f$ can be 
determined  by setting $\varepsilon =
\varepsilon_m$ in Eq.\ (\ref{eq:above}), which results in the form
\begin{eqnarray}
  \label{eq:critic}
  t_f \approx \frac{\beta\pi}{2}\sqrt{\frac{\varepsilon_m}{E}}\left(
    \sigma_{\rm o} -  
  \sigma_{c}\right)^{-1/2},
\end{eqnarray}
in accordance with the above general arguments.

A further important general property of $\varepsilon(t)$ that can be
deduced from Eqs.\ (\ref{eq:eom},\ref{eq:integ}) is that at the time
to failure $t_f$ the deformation rate $\displaystyle{d\varepsilon/dt}$
diverges. For disorder distributions $P(\varepsilon)$ defined in a
finite interval 
the exponent is universal $\displaystyle{d\varepsilon/dt
\approx (t_f-t)^{-1/2}}$.

\begin{figure}
\begin{center}
\psfrag{aa}{{\large $Et/\beta$}}
\psfrag{bb}{{\LARGE $\frac{1}{N}\frac{\partial N_b}{\partial t}$}}
\epsfig{bbllx=0,bblly=0,bburx=240,bbury=210,
file=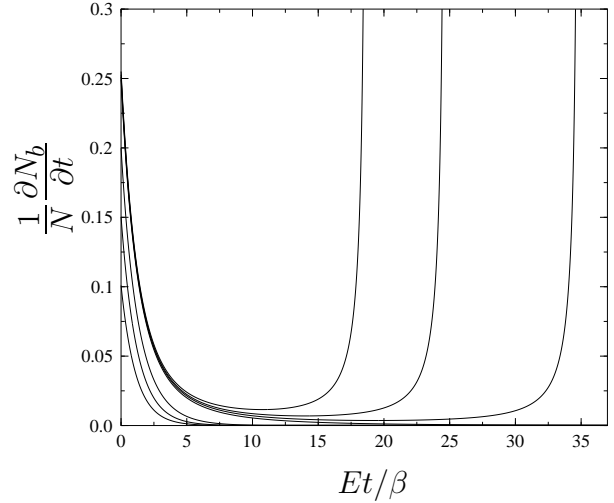, width=8.5cm}
\caption{The analytic solution for the breaking rate Eq.\
(\ref{eq:acoustic}) for several 
values of $\sigma_{\rm o}$ below and above $\sigma_{\rm c}$}
\label{fig:acoust}\end{center}
\end{figure}

In order to obtain information about the gradual breaking of fibers
during the creep process, in the experiments the acoustic
signal emitted by breaking events in a short time interval 
is investigated. 
In our fiber bundle model the number
of fibers $N_b(t)$ which have been broken up to time $t$ can be
determined as $N_b(t)=N P(\varepsilon(t))$, and hence, its
derivative provides the quantity
\begin{eqnarray}
  \label{eq:acoustic}
  \frac{1}{N}\frac{\partial N_b}{\partial t} = \frac{dP}{d
    \varepsilon}  \frac{d \varepsilon}{d t} 
   =   \frac{p(\varepsilon)E\varepsilon}{\beta}
   \left[\frac{\sigma_{\rm o}}{E\varepsilon\left[1-P(\varepsilon)\right]}   
   -1\right],
\end{eqnarray}
which is a measure for the acoustic response. The behaviour 
of Eq.\ (\ref{eq:acoustic}) for the uniform distribution is
illustrated in Fig.\ \ref{fig:acoust}, 
where it can be observed that the acoustic activity, {\it i.e.} fiber
breaking, practically disappears in the plateau region of
$\varepsilon(t)$ (compare to Fig.\ \ref{fig:eps}), however, it
diverges at $t_f$ due to the diverging 
deformation rate. 
Since during a creep test
$\varepsilon(t)$ is monitored from which ${\displaystyle
  d\varepsilon/dt}$ can be calculated, furthermore, 
$\displaystyle{\partial N_b/\partial t}$ is measured by means of
acoustic emission techniques, Eq.\
(\ref{eq:acoustic}) makes possible to determine
experimentally the 
distribution of the failure thresholds $p(\varepsilon_d)$. 

\begin{figure}[htb]
\begin{center}
\epsfig{bbllx=70,bblly=340,bburx=425,bbury=650,
file=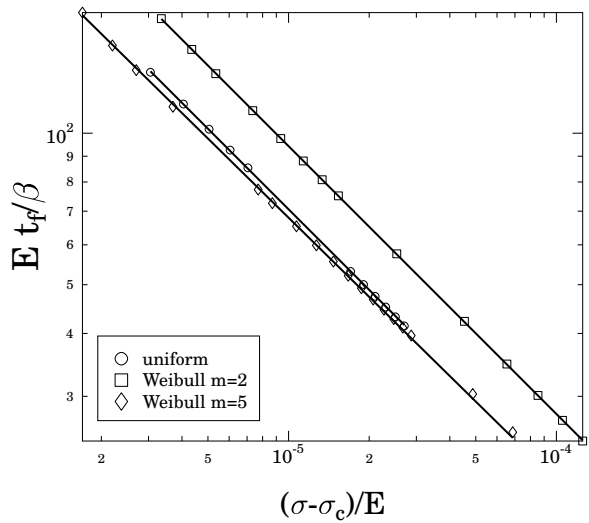, width=8.5cm}
\caption{The behaviour of the time to failure $t_f$ for uniform
and Weibull distributions with two different Weibull moduli for the
GLS case. All the
three curves are parallel to each other on a double logarithmic plot
with an exponent close to 0.5 in agreement with the general result
Eq.\ (\ref{critical}).
}
\label{fig:simcrit}\end{center}
\end{figure}

To complement  the predictions of the analytic approach Monte Carlo
simulations of the failure process have been performed using 
global load sharing (GLS) and local load sharing (LLS) for the stress
redistribution. The GLS simulation of the creep failure process of a
bundle of $N$ fibers proceeds as follows: $(i)$ random breaking thresholds
$\varepsilon_{i}, \ i=1, \ldots , N$ were chosen according to a
probability distribution $p$, then the thresholds 
were put into increasing order. $(ii)$ Since the fibers break one-by-one, 
the actual load 
on the fibers after the failure of $i$ fibers is $\sigma_i = \sigma_o
N/(N-i)$ where $i=0, \ldots , N-1$, 
and the time between the breaking of the $i$th and $i+1$th fibers
reads as ${\displaystyle   t_i = -\frac{\beta}{E} \ln \left[
  \left(\varepsilon_{i+1}-\frac{\sigma_i}{E}\right)/
\left(\varepsilon_{i}-\frac{\sigma_i}{E}\right) \right].}  $
$(iii)$ Finally, the time as a function of $\varepsilon$ can be
obtained as $t(\varepsilon_{i}) = \sum_{j=0}^{i}
t_j(\varepsilon_{j})$ from which
$\varepsilon_{i}(t)$ can be determined.  
The time to failure $t_f$ of a finite bundle is defined as the time of 
the failure of the last fiber. To test the validity of the power law
behaviour of $t_f$ given by Eq.\ (\ref{critical}) simulations were
performed with various distributions in the framework of GLS. The
results are presented in Fig.\ \ref{fig:simcrit} where an excellent
agreement of the simulations and the analytic results can be
observed. 
The macroscopic strain of the system $\varepsilon(t)$ and
the acoustic  response  obtained by simulations was also found to be
in agreement with the analytic results. 

\begin{figure}[htb]
\begin{center}
\epsfig{bbllx=140,bblly=340,bburx=475,bbury=640,
file=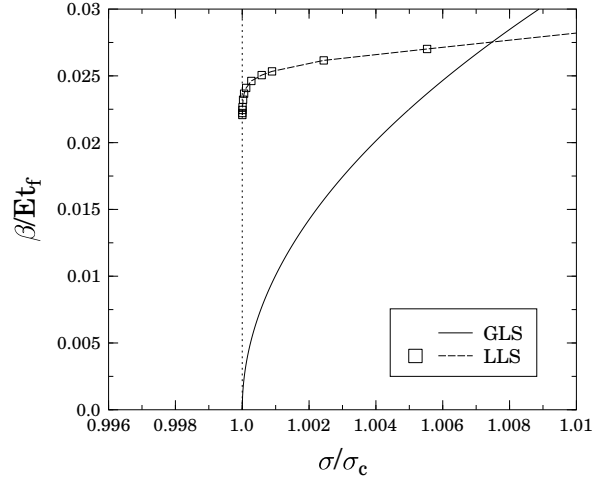, width=7.5cm}
\caption{Comparison of $1/t_f$ for the LLS and GLS cases as a function 
  of the external load.
}
\label{fig:compare}\end{center}
\end{figure}

To clarify how the damage process and the behaviour of $t_f$
is affected by the range of interaction among fibers, {\it i.e.} by the
range of load sharing we performed simulations with LLS
on a square lattice of $200 \times 200$ sites, redistributing the load
of the failed fiber on its nearest neighbors. 
The critical load $\sigma_c$ was first determined as the
static fracture strength of a dry fiber bundle with LLS assuming perfectly
elastic behaviour for the fibers. Comparing the results of the LLS
simulations to the global load sharing results 
it was observed that
above $\sigma_c$ the failure  of the viscoelastic bundle occurs much
more abruptly than in the case  
of GLS.     
Varying $\sigma$ as a control parameter the two regimes of the creep
rupture process are characterized by an infinite life time below
$\sigma_c$ and by a finite one above the critical point.
The nature of the transition between the regimes in the global and
local load sharing models can be characterized by studying  $1/t_f$ 
as a function of the control parameter $\sigma$.
In Fig.\ \ref{fig:compare} it can be observed that below the critical
point, when no global failure occurs,  $1/t_f$ is zero, while above
$\sigma_c$ it takes a finite value for both LLS and GLS. However, the
behaviour of   $1/t_f$ in the vicinity of $\sigma_c$ is completely
different in the two cases, for 
GLS the transition is continuous, while for LLS $1/t_f$ has a finite
jump, analogously to a second and first order phase transition,
respectively. 

Summarizing, we studied the creep rupture of viscoelastic fiber
composites by enhancing the classical fiber bundle model. Varying the
external load, two regimes
of the creep process were revealed characterized by an infinite life
time below, and by a finite one above the critical load. The
transition between the two regimes was found to be continuous for
global load sharing, while it is abrupt for localized load
redistribution, analogous to a second and first order phase
transition, respectively.

We are grateful to S.\ Aicher, G.\ Dill-Langer, F.\ Wittel, F.\ Tzchichholz, and to
J. Astr\"om for 
valuable discussions. 
This work was supported by the project SFB381.
F.\ Kun  acknowledges financial support of
the Alexander von Humboldt Foundation
(Roman Herzog Fellowship) and of the B\'olyai J\'anos Fellowship of
the Hungarian Academy of Sciences.

\end{multicols}


\begin{thebibliography}{100}

\bibitem{laufenberg}T.\ L.\ Laufenberg, L.\ C.\ Palka, and J. Dobbin
  McNatt, {\it Creep and Creep-Rupture Behaviour of Wood-Based
    Structural Panels}, preprint 1999.

\bibitem{pu} J.\ H.\ Pu, R.\ C.\ Tang, W.\ C.\ Davis, Forest Product
  Journal {\bf 42}, 49 (1994).
\bibitem{chiao} T.\ T.\ Chiao, C.\ C.\ Chiao, and R.\ J.\ Sherry,
  Fracture Mechanics and Technology {\bf 1}, 257 (1977).

\bibitem{otani} H.\ Otani, S.\ L.\ Phoenix, and P.\ Petrina, J.\
  Mater.\ Sci.\ {\bf 26}, 1955 (1991)

\bibitem{farq} D.\ S.\ Farquhar, F.\ M.\ Mutselle, S.\ L.\ Phoenix and 
  R.\ L.\ Smith, J.\ Mater.\ Sci.\ {\bf 24}, 2151 (1989).



\bibitem{phoenix} M.\ Ibnabdeljalil and S.\ L.\ Phoenix, J.\
  Mech. Phys. Solids {\bf 4}, 897 (1995). 

\bibitem{coleman} B.\ D.\ Coleman, J.\ Appl.\ Phys.\ {\bf 29}, 968
  (1958). 

\bibitem{gerhards} C.\ C.\ Gerhards and C.\ L.\ Link, Wood and Fiber
  Science {\bf 19}, 147 {1987}.

\bibitem{daniels} H.\ E.\ Daniels, Proc.\ R.\ Soc.\ London {\bf A
    183}, 405 (1945). 

\bibitem{hansen} M.\ Kloster, A.\ Hansen, and P.\ C.\ Hemmer, Phys.\
  Rev.\ E {\bf 56}, 2615 (1997).

\bibitem{phoenix1} D.\ G.\ Harlow and S.\ L.\ Phoenix, J.\ Composite
  Mater. {\bf 12}, 195 (1978).

\bibitem{leath}P.\ L.\ Leath and P.\ M.\ Duxbury, Phys.\ Rev.\ B {\bf
    49}, 14905 (1994).

\bibitem{zhang} S.\ D.\ Zhang and E.\ J.\ Ding, Phys.\ Rev.\ B {\bf
    53}, 646  (1996). 
\bibitem{delapl} A.\ Delaplace, S.\ Roux, and G.\ Pijaudier-Cabot,
  Int.\  J.\ Solids Struct.\ {\bf 36}, 1403 (1999).

\bibitem{sornette1}D.\ Sornette, J.\ Phys. {\bf A 22}, L243 (1989).
\bibitem{sornette2}D.\ Sornette, J.\ Phys.\ France {\bf 50}, 745
  (1989). 
\bibitem{sornette3}J.\ V.\ Andersen, D.\ Sornette, and K.-T.\ Leung,
  Phys.\ Rev.\ Lett.\ {\bf 78}, 2140 (1997).
\bibitem{kun2}
        F.\ Kun, S.\ Zapperi, and H.\ J.\ Herrmann, 
       European Physical Journal B{\bf 17}, 269 (2000). 
\bibitem{yamir} Y.\ Moreno, J.\ B.\ Gomez, A.\ F.\ Pacheco, 
       Phys.\ Rev.\ Lett.\ {\bf 85}, 2865 (2000). 
\bibitem{menezes} I.\ L.\ Menezes-Sobrinho, A.\ T.\ Bernardes, and
  J.\ G.\ Moreira, Phys.\ Rev.\ E {\bf 63} 25104(R) (2001). 

\bibitem{acoust} T.\ F.\ Drouillard and F.\ C.\ Beall, J.\ Acoustic
  Emission {\bf 9}, 215 (1990). 
\end{thebibliography}
\end{document}